\documentclass[usenatbib]{mn2e}
\usepackage{amssymb,amsmath,graphicx}

\begin{document}
\title{New biorthogonal potential--density basis functions}
\author[A. Rahmati and M.A. Jalali]
  {Alireza~Rahmati$^1$\thanks{rahmati@strw.leidenuniv.nl (AR)} and
  Mir Abbas~Jalali$^{2,3}$\thanks{mjalali@sharif.edu (MAJ)} \\
  $^1$Sterrewacht Leiden, Leiden University, P.O. Box 9513, 2300 RA Leiden, 
  The Netherlands\\
  $^2$Sharif University of Technology, Azadi Avenue, Tehran, Iran \\
  $^3$School of Astronomy, Institute for Studies in Theoretical Physics 
  and Mathematics (IPM), P.O. Box 19395-5531, Tehran, Iran}
  
\maketitle

\begin{abstract}
We use the weighted integral form of spherical Bessel functions, 
and introduce a new analytical set of complete and biorthogonal 
potential--density basis functions. The potential and density 
functions of the new set have finite central values and they 
fall off, respectively, similar to $r^{-(1+l)}$ and $r^{-(4+l)}$ 
at large radii where $l$ is the latitudinal quantum number of 
spherical harmonics. The lowest order term associated with $l=0$ 
is the perfect sphere of de Zeeuw. Our basis functions are 
intrinsically suitable for the modeling of three dimensional, 
soft-centred stellar systems and they complement the basis sets 
of Clutton-Brock, Hernquist \& Ostriker and Zhao. We test the 
performance of our functions by expanding the density and 
potential profiles of some spherical and oblate galaxy models.
\end{abstract}



\section{Introduction}
Solving Poisson's equation is an important step in the study 
of self-gravitating stellar systems \citep{BT08}. Expanding 
the density distribution and its conjugate potential field in 
terms of a complete basis set is one of the most efficient methods 
that investigators have extensively applied to $N$-body simulations 
\citep{FP84,HO92,ES95,MZ97,WK07,Buyle07} and the first-order 
stability analysis of both flat \citep{K77,PC97,JH05,J07} and 
three dimensional galaxies \citep{Saha91,W91}. Consequently, 
the success of those studies highly depends on the choice of 
basis set. Desirable potential and density basis functions 
should be biorthogonal and converge rapidly in order to 
decrease the computational noise and cost. Nevertheless, finding 
a suitable basis set is not an easy task and only few analytical 
basis sets have been found for three dimensional stellar systems. 

For stellar systems of finite size, spherical Bessel functions are the 
classical biorthogonal eigenfunctions of the Laplace operator and they 
have been used in the stability analysis of certain spherical galaxies 
\citep{FP84,All90,W91}. For galaxy models of infinite extent, three 
biorthogonal potential--density (PD) basis sets have been developed 
by Clutton-Brock (1973, hereafter CB73), 
Hernquist \& Ostriker (1992, hereafter HO92) and \citet{Zhao96}.
CB73 and HO92 set the lowest order terms of their basis functions 
to the \citet{P11} and \citet{H90} models while \citet{Zhao96} 
uses an $\alpha$-model with the density
\begin{equation}
\rho(r)=\frac{C}{r^{2-1/\alpha}\left ( 1+r^{1/\alpha}
\right )^{2+\alpha}},
\end{equation}
where $C$ is a constant parameter. The lowest order term of a PD 
set does not necessarily need to be spherical \citep{Syer95}, 
but an orthonormalization using the standard Gram-Schmidt 
procedure must be adopted \citep{Saha91,Robijn96} to guarantee 
the completeness of the set.  

Apart from the quoted analytic basis functions, numerically 
generated sets have also become available. \citet{W99} assumed 
the form of the lowest order basis functions and numerically 
solved the Strum-Liouville equation to obtain biorthogonal 
basis functions of higher orders. Despite this worthwhile contribution, 
the propagation of computational noise during the application of 
numerical basis functions has become problematic in recent $N$-body 
experiments \citep{KCE08}, which justify the ongoing search for 
new analytical basis functions.

In this paper we introduce a new analytical set of biorthogonal
PD basis functions whose potential and density components have 
finite central values, fall off similar to HO92 functions as 
$r\rightarrow \infty$, and their lowest order term is the 
perfect sphere of \citet{dZ85}. We derive and evaluate the 
weighted integral forms of spherical Bessel functions in 
\S\ref{sec:potential-density-pairs}, and obtain the radial 
basis functions in terms of associated Legendre functions. 
In \S\ref{sec:modeling-galaxy-profiles}, we use the new basis 
set and generate the series representations of certain spherical 
and oblate galaxy models. We end the paper with concluding
remarks.

\section{Potential--density pairs}
\label{sec:potential-density-pairs}

We define $\mathbf{r}=\left ( r,\theta,\phi \right )$ as the 
position vector expressed in terms of usual spherical coordinates, 
with $r$, $\theta$ and $\phi$ being the radial distance from the 
origin, co-latitude and azimuthal angle, respectively. We also  
assume that the mean-field potential and density functions of
a stellar system admit the following expansions
\begin{subequations}
\label{eq:expand-pot-den}
\begin{align}
\Phi(\mathbf{r}) &= \sum_{nlm}P_{nlm} \Phi_{nlm}(\mathbf{r}),
\label{eq:expand-potential}\\
\rho(\mathbf{r}) &= \sum_{nlm}D_{nlm} \rho_{nlm}(\mathbf{r}).
\label{eq:expand-density}
\end{align}
\end{subequations}
The basis functions $\Phi_{nlm}(\mathbf{r})$ and 
$\rho_{nlm}(\mathbf{r})$ satisfy Poisson's equation
\begin{equation}
\nabla^2 \Phi_{nlm}(\mathbf{r}) = 
4 \pi G \rho_{nlm}(\mathbf{r}),
\label{eq:3d-poisson-equation}
\end{equation}
where $n$ and $l$ are the radial and latitudinal quantum numbers
corresponding to $r$ and $\theta$, respectively, and $m$ is the 
azimuthal Fourier number associated with $\phi$. $G$ is the 
universal constant of gravitation. We proceed with a case that 
$\rho_{nlm}$ is proportional to $\Phi_{nlm}$. This reduces Poisson's 
equation to the eigenvalue problem
\begin{equation}
\nabla^2 \Phi_{nlm}(\mathbf{r}) = 
-4\pi G k^2 \Phi_{nlm}(\mathbf{r}),
\label{eq:eigenvalue-equation}
\end{equation}
that involves the Laplace operator $\nabla ^2$ and a constant
parameter $k$. Since the Laplace operator is Hermitian, its
associated eigenfunctions form a complete biorthogonal basis set.
The coefficients $P_{nlm}$ and $D_{nlm}$ thus become identical. The
spherical harmonics $Y_{lm}(\theta,\phi)$ and the spherical Bessel
functions $j_n(kr)$ are the classical solutions of
(\ref{eq:eigenvalue-equation}). 

While $Y_{lm}(\theta,\phi)$ show an acceptable performance in 
the expansion of physical quantities in terms of angle variables, 
Bessel functions do not look like galactic profiles and can not 
generate efficient expansions \citep{W99}. We extend the method
of \citet{CB72} to three dimensional systems and express the 
eigenfunctions as
\begin{subequations}
\label{eq:pot-den-pair}
\begin{align}
\Phi_{nlm}(\mathbf{r}) &= -Y_{lm}(\theta,\phi)\psi_{nl}(r),
\label{eq:potential-basis} \\
\rho_{nlm}(\mathbf{r}) &= Y_{lm}(\theta,\phi)\rho_{nl}(r),
\label{eq:density-basis}
\end{align}
\end{subequations}
where 
\begin{subequations}
\label{eq:radial-basis-1}
\begin{align}
\psi_{nl}(r) &= \int_{0}^{\infty}j_{l}(kr)g_{nl}(k)dk,
\label{eq:radial-potential-basis-1} \\
\rho_{nl}(r) &= \frac{1}{4 \pi G} 
\int_{0}^{\infty}j_{l}(kr)g_{nl}(k)k^2dk.
\label{eq:radial-density-basis-1}
\end{align}
\end{subequations}
The functions $g_{nl}(k)$ ($n,l=0,1,2,\cdots$) are to-be-determined 
functions that we require to satisfy the biorthogonality condition
\begin{equation}
\int \Phi_{nlm}(\mathbf{r})[\rho_{n'l'm'}(\mathbf{r})]^*d\mathbf{r} = 
I_{nlm}\delta_{nn'}\delta_{ll'}\delta_{mm'}.
\label{eq:biorthogonality-basic}
\end{equation}
Here, the asterisk denotes complex conjugation and $\delta_{ii'}$ is
the Kronecker delta. Substituting from (\ref{eq:pot-den-pair})
and (\ref{eq:radial-basis-1}) in (\ref{eq:biorthogonality-basic}) and 
using the identity
\begin{equation}
\int_{-1}^{1} d \left ( \cos \theta \right )
\int_{0}^{2\pi} d\phi Y_{lm}(\theta,\phi)
Y^*_{l'm'}(\theta,\phi)=\delta_{ll'}\delta_{mm'},
\end{equation}
the orthogonality condition (\ref{eq:biorthogonality-basic}) reduces to
\begin{eqnarray}
&{}& -\frac{1}{4\pi G}\int_{0}^{\infty}
g_{nl}(k) dk \int_{0}^{\infty} g_{n'l}
\left ( k'\right ) k'^2 dk' \nonumber \\
&{}& \qquad \times \int_{0}^{\infty} j_l(kr) j_{l}\left (k'r \right )
r^2 dr =I_{nlm}\delta_{nn'}.
\label{eq:reduced-orthogonality}
\end{eqnarray}
The innermost integral on the left-hand side of
(\ref{eq:reduced-orthogonality}) is evaluated according to the
Fourier-Bessel theorem \citep{U72} as
\begin{equation}
\int_{0}^{\infty}j_{l}(kr)j_{l}(k'r)r^2dr=
\frac{\pi}{2k^2}\delta(k'-k),
\label{eq:Fourier-Bessel-theorem}
\end{equation}
with $\delta(k'-k)$ being the Dirac delta function. Substituting
(\ref{eq:Fourier-Bessel-theorem}) in (\ref{eq:reduced-orthogonality})
leads to
\begin{equation}
-\frac{1}{8G}\int_{0}^{\infty}
g_{nl}(k)g_{n'l}(k)dk = \overline I_{nl}\delta_{nn'} \equiv I_{nlm}\delta_{nn'}.
\label{eq:orthogonality-of-g-function}
\end{equation}
This condition requires $g_{nl}(k)$ to be any orthogonal set of
functions over the semi-infinite $k$-domain. Our special choice
is $g_{nl}(k)= k^{l}L_{n}^{2l}(2k)e^{-k}$ where $L^{p}_q(k)$
are the associated Laguerre polynomials that obey the following 
orthogonality relation
\begin{equation}
\int_{0}^{\infty}e^{-k}k^{p}L_{q}^{p}(k)L_{q'}^{p}(k)dk=
\frac{(q+p)!}{q!}\delta_{qq'}.
\end{equation}
Consequently, the constant parameters on the right-hand
side of equation (\ref{eq:orthogonality-of-g-function}) become
\begin{equation}\label{bi-cons}
\overline I_{nl} = -\frac{(n+2l)!}{G 2^{2l+4} n!},
\end{equation}
and our radial basis functions read
\begin{subequations}
\label{eq:radial-basis-functions-2}
\begin{align}
\psi_{nl}(r)&=\int_{0}^{\infty}j_{l}(kr)L_{n}^{2l}(2k)e^{-k}k^{l}dk,
\label{eq:radial-potential-basis-2}\\
\rho_{nl}(r)&=\frac{1}{4 \pi
G}\int_{0}^{\infty}j_{l}(kr)L_{n}^{2l}(2k)e^{-k}k^{l+2}dk.
\label{eq:radial-density-function-2}
\end{align}
\end{subequations}
\begin{figure*}
\centerline{\hbox{\includegraphics[width=0.5\textwidth]
             {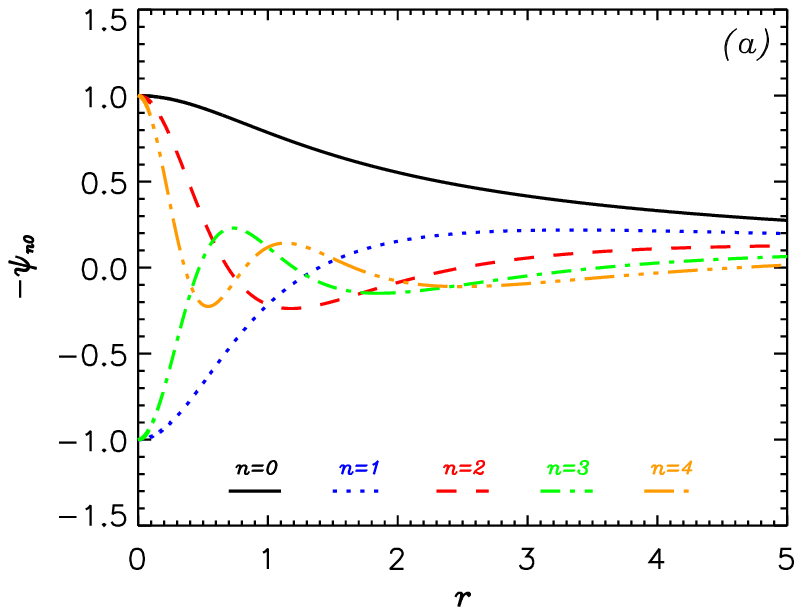}}
            \hbox{\includegraphics[width=0.5\textwidth]
             {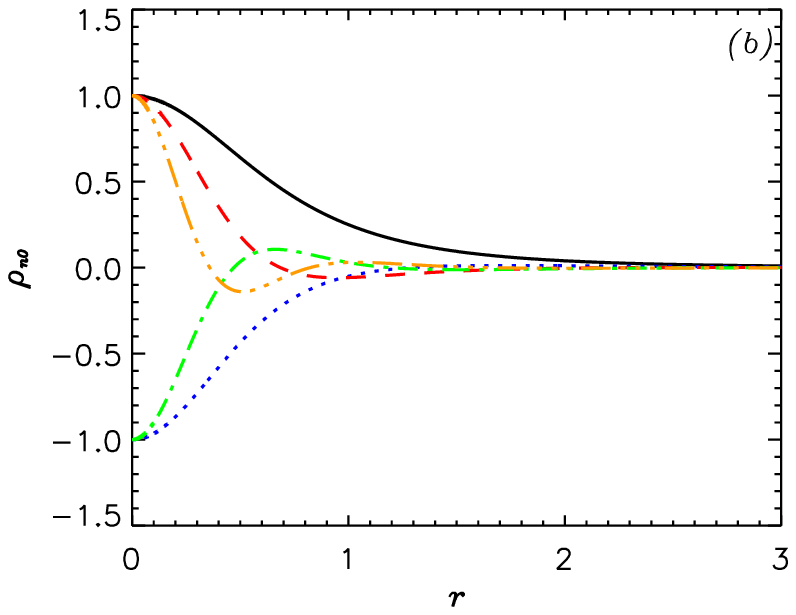}} }
\centerline{\hbox{\includegraphics[width=0.5\textwidth]
             {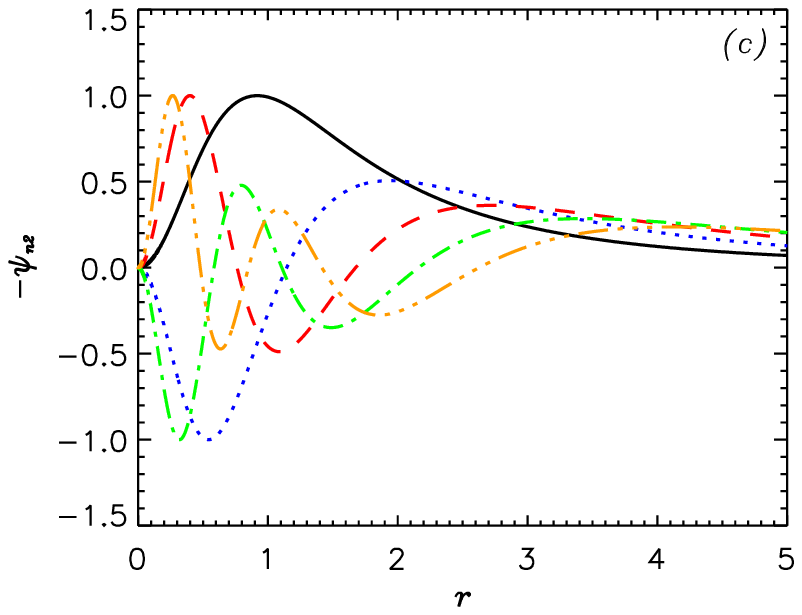}}
            \hbox{\includegraphics[width=0.5\textwidth]
             {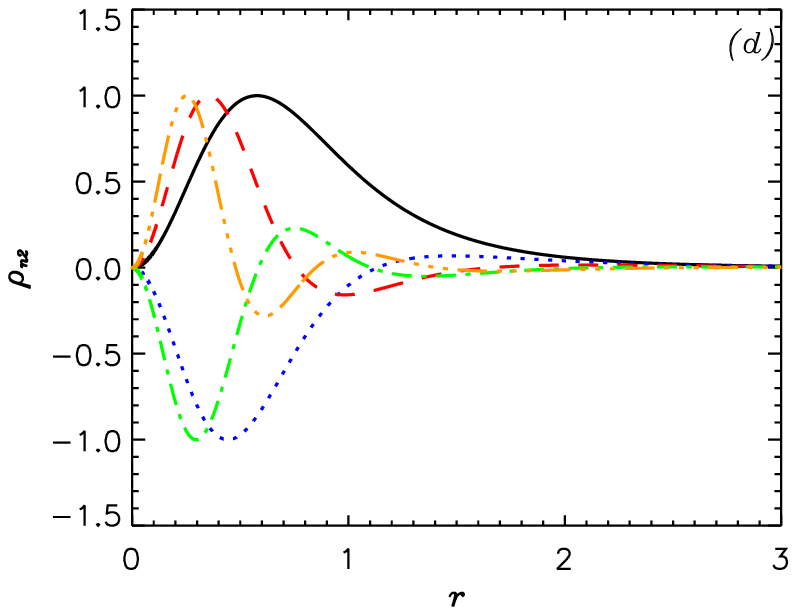}} }
\caption{Left panels display the radial parts of the potential basis
functions, $\psi_{nl}(r)$, for $n=0,1,2,3,4$, and right panels show
their conjugate density functions $\rho_{nl}(r)$. Top and bottom
panels correspond to $l=0$ and $l=2$, respectively. All functions
have been normalised to their maximum values.
} \label{fig1}
\end{figure*}

The integrals in (\ref{eq:radial-basis-functions-2}) converge rapidly,
which makes their evaluation a straightforward task by numerical methods.
However, closed-form analytical expressions can also be derived for
$\psi_{nl}(r)$ and $\rho_{nl}(r)$ as we explain below. We utilise the
series form of the Laguerre functions
\begin{equation}\label{ap2}
L_{n}^{2l}(2k)=\sum_{i=0}^n(-1)^i
\begin{pmatrix} n+2l\\
                n-i
\end{pmatrix} \frac{(2k)^i}{i!},
\end{equation}
and express $j_l(kr)$ in terms of Bessel functions to rewrite
(\ref{eq:radial-density-function-2}) in the form
\begin{eqnarray}
&{}& \rho_{nl}(r) = \frac{1}{4 \pi G} \sum_{i=0}^n(-2)^i
\begin{pmatrix} n+2l \\
                n-i
\end{pmatrix}
\sqrt{\frac{\pi}{2r}}\frac{1}{i!} \nonumber \\
&{}& \qquad \qquad \qquad \times
\int_{0}^{\infty}J_{l+\frac{1}{2}}(kr)e^{-k}k^{l+i+3/2}dk.
\label{eq:rho-after-first-substitute}
\end{eqnarray}
Carrying out a change of independent variable as $kr\rightarrow u$,
transforms equation (\ref{eq:rho-after-first-substitute}) to
\begin{eqnarray}
&{}& \rho_{nl}(r)=\frac{1}{4 \pi G}
\sum_{i=0}^n(-2)^i
\begin{pmatrix} n+2l \\
                n-i
\end{pmatrix}
\sqrt{\frac{\pi}{2}}\frac{r^{-l-i-3}}{i!} \nonumber \\
&{}& \qquad \qquad \qquad \times 
\int_{0}^{\infty}J_{l+\frac{1}{2}}(u)
e^{-u/r} u^{l+i+3/2}du.
\label{eq:rho-after-second-substitute}
\end{eqnarray}
The integral in (\ref{eq:rho-after-second-substitute}) can be 
calculated using equation (6.621) in \citet{GR00}. Defining 
$\xi=1/\sqrt{1+r^2}$, $\nu_1 = l+i+3$, 
$\nu_2=l+i+3/2$ and $\mu=-1/2-l$, we obtain
\begin{eqnarray}
&{}& \rho_{nl}(r) =\frac{1}{4 \sqrt{2\pi} G}
\sum_{i=0}^{n}
\frac{(2l+i+2)(2l+i+1)(n+2l)!}{i!(n-i)!} \nonumber \\
&{}& \qquad \qquad \qquad \times (-2)^i \xi^{\nu_1} 
\left (1-\xi^2 \right )^{-1/4} 
P^{\mu}_{\nu_2} \left ( \xi \right ),
\label{eq:final-closed-form-rho}
\end{eqnarray}
where $P^{\mu}_{\nu}(x)$ are associated Legendre functions. 
Following a similar procedure, one can show that
\begin{equation}
\psi_{nl}(r)=\sqrt{ \frac{\pi}{2} }
\sum_{i=0}^{n} \frac{(n+2l)!}{i!(n-i)!}
(-2)^i \xi^{\nu_3}
\left (1-\xi^2\right )^{-1/4} P^{\mu}_{\nu_4}
\left (\xi \right ),
\label{eq:final-closed-form-psi}
\end{equation}
where $\nu_3=l+i+1$ and $\nu_4=l+i-1/2$. 
The associated Legendre functions can be determined 
through the recursive relations \citep{GR00}
\begin{subequations}
\label{eq:recursion}
\begin{align}
& (\nu-\mu+1)P_{\nu+1}^{\mu}(\xi)+(\nu+\mu)P_{\nu-1}^{\mu}(\xi) =
(2\nu+1)\xi P_{\nu}^{\mu}(\xi), \\
& P_{\nu-1}^{\mu}(\xi) - P_{\nu+1}^{\mu}(\xi)
= (2\nu+1)(1-\xi^2)^{1/2} P_{\nu}^{\mu-1}(\xi),
\end{align}
\end{subequations}
which start from
\begin{equation}
P_{\nu}^{-1/2} \left (\xi \right )
= \sqrt{\frac{2}{\pi \sin \beta} }
\frac{\sin \left [\left (\nu+1/2 \right )\beta \right ]}
{\left (\nu+1/2 \right ) },~~\cos\beta=\xi.
\label{eq:starting-functions}
\end{equation}

The lowest order members of our PD family are
\begin{subequations}
\label{eq:new-spherical-galaxy-model}
\begin{align}
\psi_{00}(r) &= -\frac {1}{r} \arctan r, \label{eq:new-pot-00}\\
\rho_{00}(r) &= \frac{1}{2\pi G}\frac{1}{\left (1+r^2 \right )^2},
\label{eq:new-density-00}
\end{align}
\end{subequations}
which define the perfect sphere of \citet{dZ85}. We have therefore
found a biorthogonal basis set that is distinct from CB73, HO92 and 
Zhao's (1996) functions. Moreover, from (\ref{eq:starting-functions}) 
and the recursive relations (\ref{eq:recursion}), we deduce that the 
functions $P^{\mu}_{\nu_4}(\xi)$ and $P^{\mu}_{\nu_2}(\xi)$ behave,
respectively, similar to $r^0$ and $r^{-1}$ in the limit of 
$r\rightarrow \infty$. It can thus be verified that 
$\psi_{nl}(r)\sim r^{-(1+l)}$ and $\rho_{nl}(r)\sim r^{-(4+l)}$ 
hold at large radii. The potential functions of CB73, HO92 and ours 
have finite central values and they fall off similar to $r^{-(1+l)}$ 
at large radii. Our density functions are analytic at the galactic 
centre as are the functions of CB73, but they behave like HO92 
functions in the limit of $r\rightarrow \infty$. The best performance 
of our basis set is thus expected in soft-centred systems whose outer 
density profiles are similar to $r^{-4}$.

In Figure \ref{fig1}, we have displayed several members 
of our basis functions for $l=0,2$. At the centre, both the
potential and density functions have finite, non-zero values 
for $l=0$, and they vanish there for $l\neq0$. The expected yet
interesting property of $\psi_{nl}(r)$ and $\rho_{nl}(r)$ is 
their oscillatory nature. The number of peaks of our functions 
(in the radial direction) is equal to $n+1$. Our numerical 
experiments show that the series built by oscillatory functions 
have a faster and more accurate mean-convergence compared to 
functions that do not share this feature.

Our functions have a length scale that has been set to unity so
far. In general, changing the length scale is necessary
to reconstruct galaxies of different core radii. A scaling parameter
$r_0$ can be easily introduced to our formulation through replacing
$g_{nl}(k)$ with $g_{nl}(k r_0)$ \citep{CB72}. This implies the
following transformations
\begin{subequations}
\begin{align}
\Phi_{nlm}\left ( r,\theta,\phi \right ) & \rightarrow
r_0^{-1} \Phi_{nlm} \left ( r/r_0,\theta,\phi \right ), \\
\rho_{nlm}\left ( r,\theta,\phi \right )&  \rightarrow 
r^{-3}_0 \rho_{nlm} \left ( r/r_0,\theta,\phi \right ), \\
\overline I_{nl} & \rightarrow r_0^{-1} \overline I_{nl}.
\end{align}
\end{subequations}

\section{RECONSTRUCTION OF MODEL GALAXIES}
\label{sec:modeling-galaxy-profiles}

Bi-orthogonal basis functions, similar to ours, have the advantage 
that the coefficients $P_{nlm}=D_{nlm}$ in (\ref{eq:expand-pot-den}) 
can be determined using either the potential $\Phi(\mathbf{r})$ or 
the density $\rho(\mathbf{r})$ through the following formulae
\begin{eqnarray}
P_{nlm}\equiv D_{nlm} \!\! &=& \!\! \frac{1}{\overline I_{nl}}
\int \rho(\mathbf{r})[\Phi_{nlm}(\mathbf{r})]^* d\mathbf{r} \nonumber \\
\!\! &=& \!\! \frac{1}{\overline I_{nl}} 
\int \Phi(\mathbf{r})[\rho_{nlm}(\mathbf{r})]^* d\mathbf{r},
\label{eq:coefficients-determine-from-rho}
\end{eqnarray}
where we have used the orthogonality conditions (\ref{eq:biorthogonality-basic})
and (\ref{eq:orthogonality-of-g-function}). In what follows, we examine the
performance of our basis functions by the series reconstruction of
the density profiles and potential fields of some model galaxies.

\subsection{Spherical Models}
\label{sec:spherical-models}

We followed the standard procedure of using spherical harmonics
for the expansions of physical quantities in terms of angular 
variables, and introduced a new set of radial basis functions. 
So we need to examine the performance of our radial set by 
reproducing some spherical models. As case studies, we choose 
the isochrone and Plummer models of total mass $M$ and length 
scale $b$ \citep{BT08}. Our basis functions have finite values 
at the centre and it would be interesting to learn whether 
they are suitable for the reconstruction of models with 
central density cusps. For doing so, we also analyse the 
performance of our basis functions by applying them to 
Dehnen's $\gamma$-models \citep{Deh93}. The density profiles 
of Dehnen's models diverge similar to $r^{-\gamma}$ in central 
regions and fall off proportional to $r^{-4}$ at large radii. 
Dehnen's models also have a length scale $b$. The model with $\gamma=0$ 
has an intrinsic core at the centre and for $\gamma=3/2$ 
a central cusp with an intermediate slope between \citet{H90} 
and \citet{Jaffe83} models is created. In our study, we choose 
two models with $\gamma=1/2$ and $\gamma=3/2$. 

\begin{figure}
\centerline{\hbox{\includegraphics[width=0.47\textwidth]
             {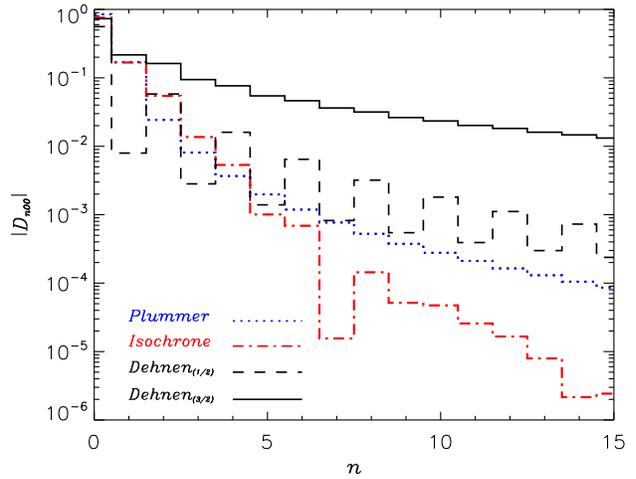}} }
\caption{The absolute magnitudes $|D_{n00}|$ of the expansion 
coefficients versus the radial quantum number $n$ for several 
spherical models.} \label{fig2}
\end{figure}

We have set $r_0=1$ and used equation (\ref{eq:coefficients-determine-from-rho}) 
to compute the coefficients of expansion $D_{nlm}$ for the Plummer, isochrone 
and Dehnen models. The parameters of the isochrone model have been set to $GM=1$ 
and $b=0.5$. For other models we have used $GM=1$ and $b=1$. Our results are 
displayed in Figure \ref{fig2}, which shows how $|D_{n00}|$ vary versus $n$. 
We note that all coefficients with $l,m\neq 0$ vanish because of spherical 
symmetry. It is evident that $|D_{n00}|$ decrease several orders of magnitude 
by including more terms in the series expansions. Although for $\gamma=1/2$ 
the coefficients of Dehnen's model fall off similar to other soft-centred 
models, they decay mildly for $\gamma=3/2$. This shows very slow and 
unfavourable convergence of our series expansion in steeper cusps as 
is expected.

\begin{figure*}
\centerline{\hbox{\includegraphics[width=0.47\textwidth]
             {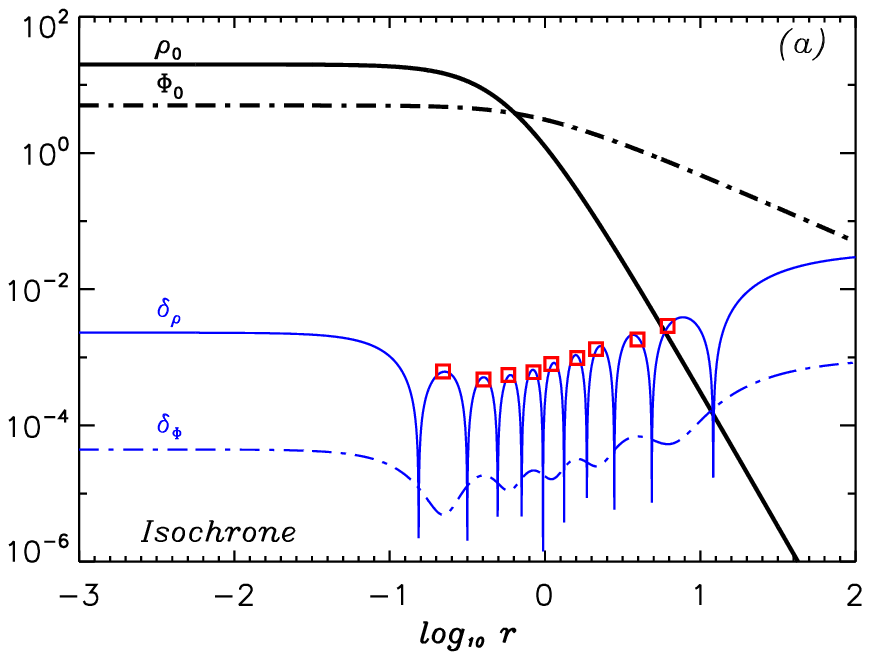}} \hspace{0.1in}
            \hbox{\includegraphics[width=0.47\textwidth]
             {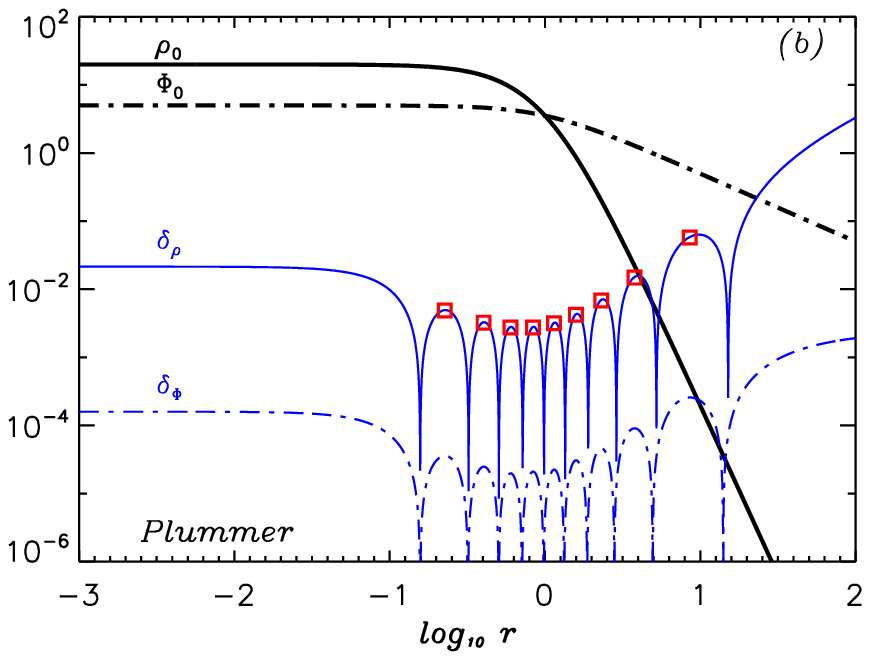}}  }
\centerline{\hbox{\includegraphics[width=0.47\textwidth]
             {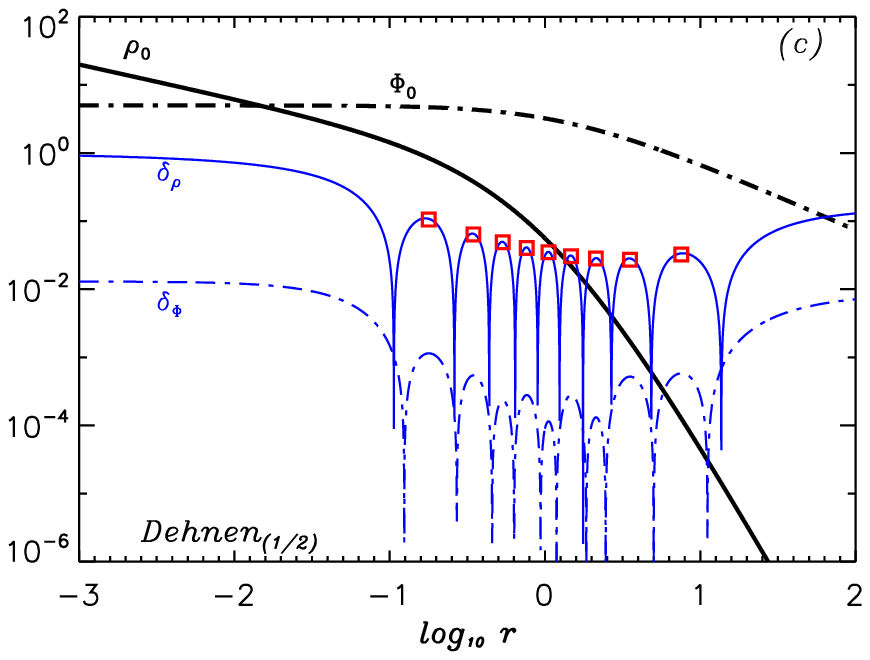}} \hspace{0.1in}
            \hbox{\includegraphics[width=0.47\textwidth]
             {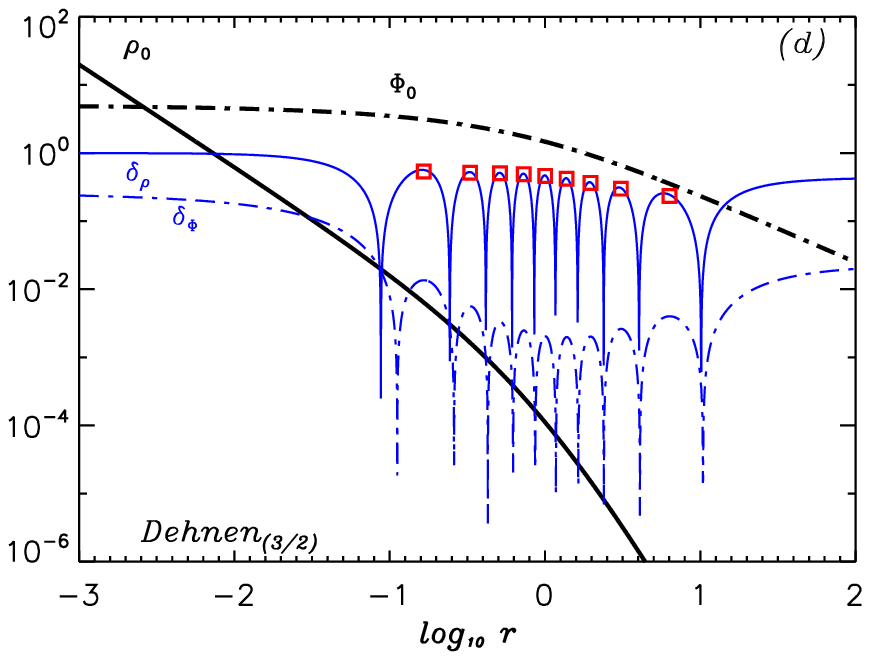}}  }	     
\caption{ The relative errors $\delta_{\Phi}$ 
(thin solid lines) and $\delta_{\rho}$ (thin dash-dotted lines) 
in the series reconstruction of the potential $|\Phi_0|$ (thick 
solid lines) and its associated density $\rho_0$ (thick dash-dotted lines). 
The basis functions are those of equations (\ref{eq:final-closed-form-rho}) 
and (\ref{eq:final-closed-form-psi}). Panels {\em a}, {\em b}, 
{\em c} and {\em d} correspond, respectively, to the isochrone, 
Plummer, Dehnen's $\gamma=1/2$ and Dehnen's $\gamma=3/2$ 
models. The functions $\Phi(r)$ and $\rho(r)$ have been computed 
by taking the first 10 basis functions ($n_{\rm max}=9$). In each 
error curve, there are $n_{\rm max}+1$ sharp minima that correspond 
to the locations of exact match between the original functions and 
their series representations. The scattered squares show the 
magnitude of $\delta_{\rho}$ calculated from equation 
(\ref{eq:relative-delta-rho-2}). The central values of $\Phi_0$
and $\rho_0$ have been normalised to some arbitrary numbers just for 
the purpose of visualising their profiles against the error curves.}  
\label{fig3}
\end{figure*}

Having expansion coefficients, the original model can be 
constructed using (\ref{eq:expand-pot-den}). Denoting the 
original PD pair by $[\Phi_0(r),\rho_0(r)]$ and their series 
representations by $[\Phi(r),\rho(r)]$, we compute the relative 
errors $E_{\rho}$=$(\rho-\rho_0)/\rho_0$ and 
$E_{\Phi}$=$(\Phi-\Phi_0)/\Phi_0$, and their absolute magnitudes 
$\delta_{\rho}=|E_{\rho}|$ and $\delta_{\Phi}=|E_{\Phi}|$ to measure 
the performance of the basis set. We have used the first 10 radial 
basis elements ($n_{\rm max}=9$) to compute $[\Phi(r),\rho(r)]$. 
The results are shown in Figure \ref{fig3}. It is seen that 
$\delta_{\Phi}$ is below 2\% in all parts of the Plummer, 
isochrone and Dehnen's $\gamma=1/2$ models, and also for 
$r>0.1$ in Dehnen's $\gamma=3/2$ model. The reason is the 
similarity of $\psi_{n0}(r)$ defined in (\ref{eq:final-closed-form-psi}) 
to the potential profiles of the chosen models. The large error 
magnitude near the centre of Dehnen's $\gamma=3/2$ model is due to 
its sharper density cusp that prohibits a simultaneous convergence 
of the density and potential series. 

The reconstruction of $\rho_0(r)$, however, has not been successful 
in Dehnen's $\gamma=3/2$ model because of its sharper cusp. Large 
values of $\delta_{\rho}$ are also observed in the central part 
of Dehnen's $\gamma=1/2$ model (due to its cuspy nature), and at 
large radii of the Plummer model due to its rapid density fall-off,
which is steeper than our $\rho_{n0}(r)\sim r^{-4}$. The isochrone 
model is the only case that has been reproduced with a reliable 
accuracy in all parts of the galaxy. In fact, the isochrone 
model shares two basic features of our new basis set: 
(i) It has a soft core. (ii) Its outer potential and density 
profiles decay, respectively, similar to $r^{-1}$ and $r^{-4}$ 
as do the envelopes of the functions $\psi_{n0}(r)$ and $\rho_{n0}(r)$. 
In Figure \ref{fig3}{\em a} for $n_{\rm max}=9$, the magnitude of 
$\delta_{\rho}$ is less than 1\% over the range $0\le r \lesssim 20$ 
and it saturates at a level of $\delta_{\rho} \approx 5\%$ for 
$r>20$. By increasing $n_{\rm max}$ to $14$, both $\delta_{\rho}$ 
and $\delta_{\Phi}$ remain smaller than $1\%$ over the range  
$0<r<100$. This result can also be deduced from Figure \ref{fig2} 
that shows a monotonic decay for $|D_{n00}|$ versus $n$. 

\begin{figure*}
\centerline{\hbox{\includegraphics[width=0.5\textwidth]
             {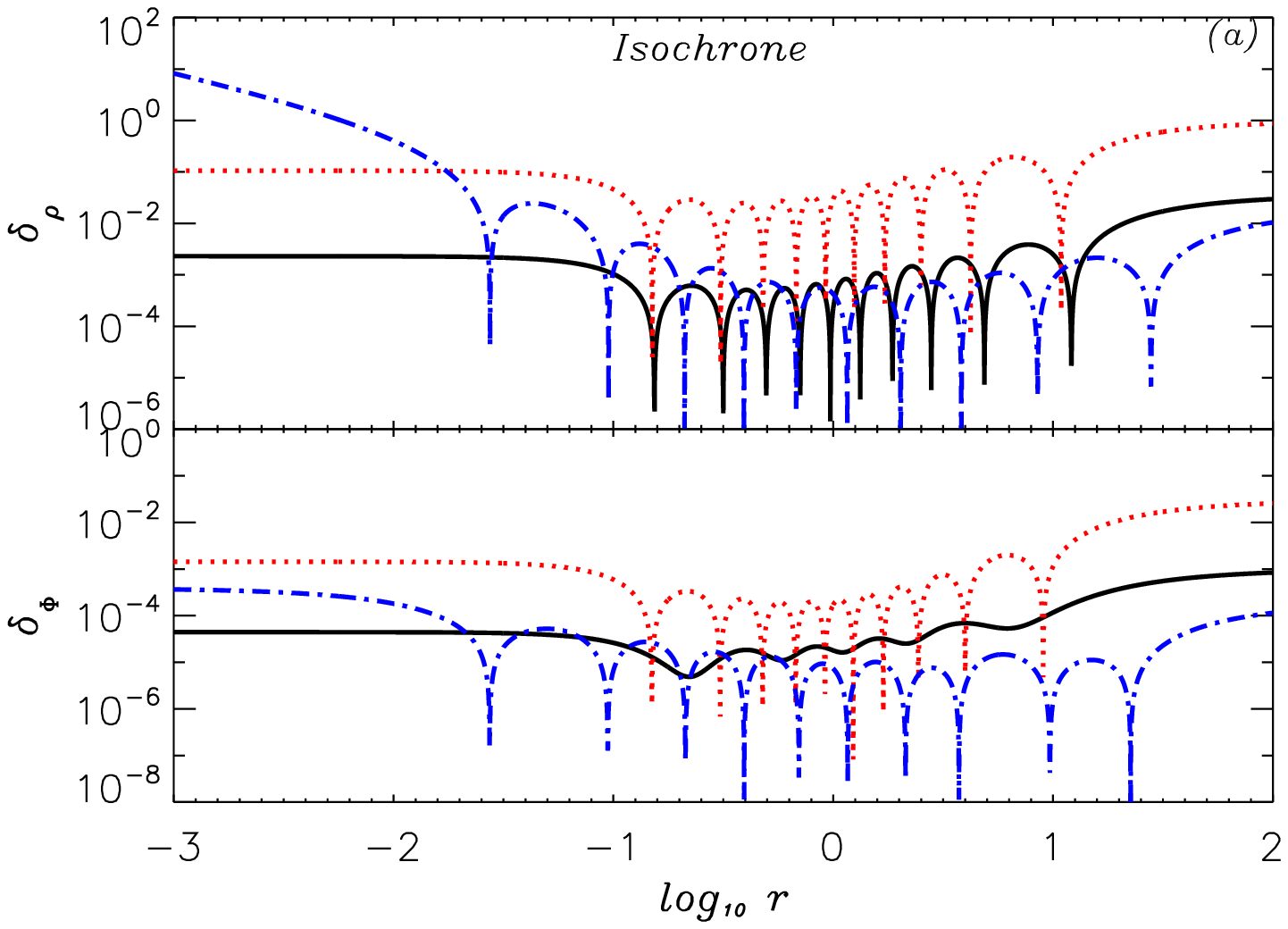}}
            \hbox{\includegraphics[width=0.5\textwidth]
             {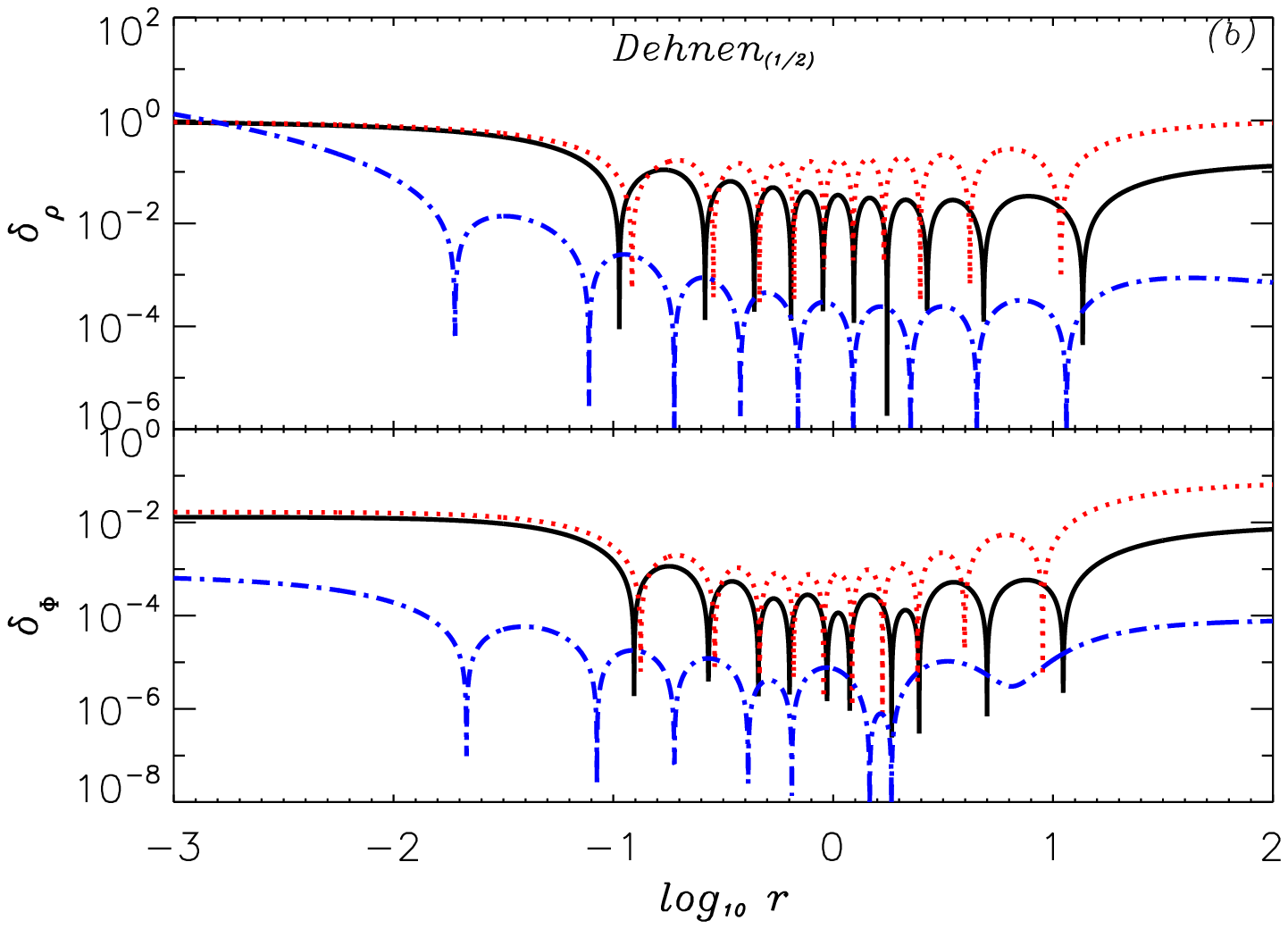}} }
\caption{Variations of $\delta_{\rho}$ ({\it top panels}) 
and $\delta_{\Phi}$ ({\it bottom panels}) for the isochrone
({\it left panels}) and Dehnen's $\gamma=1/2$ ({\it right panels}) 
models. Solid, dotted and dash-dotted lines respectively correspond 
to expansions by our, CB73 and HO92 basis functions. The length scale  
of the isochrone model is $b=0.5$ and that of Dehnen's $\gamma=1/2$
model is $b=1$. For both models we have set $GM=1$ and all basis 
functions have the length scale of $r_0=1$.} 
\label{fig4}
\end{figure*}

In general, the density error $\delta_{\rho}$ is larger than 
$\delta_{\Phi}$. We explain this by calculating $E_{\rho}$
in terms of $E_{\Phi}$ and its derivatives. The original potential 
and density functions satisfy Poisson's equation, and since our basis 
functions are biorthogonal, the relation $\nabla^2\Phi=4\pi G\rho$
also holds between the expanded quantities. We can therefore
write 
\begin{equation}
\nabla^2 \left (\Phi-\Phi_0 \right )=
4\pi G\left ( \rho-\rho_0 \right ),
\end{equation}
which is divided by $\rho_0$ to obtain 
\begin{equation}
\frac{1}{\rho_0}\nabla^2 \left (\Phi-\Phi_0 \right ) \equiv 
\frac{1}{\rho_0}\nabla^2\left ( \Phi_0 E_{\Phi} \right)
=4\pi G E_{\rho}.
\label{eq:relative-error-for-density}
\end{equation}
For the Laplace operator with spherical symmetry, equation
(\ref{eq:relative-error-for-density}) leads to   
\begin{equation}
\frac{1}{\rho_0} 
\left [ \Phi_0 \nabla^2 E_{\Phi}+E_{\Phi}\nabla^2 
\Phi_0 +2\frac{ d \Phi_0}{d r}
         \frac{ d E_{\Phi}}{d r} 
\right ] =
4\pi G E_{\rho}. \label{eq:relative-delta-rho-0}
\end{equation}
Substituting $4\pi G\rho_0$ for $\nabla^2 \Phi_0$ in 
(\ref{eq:relative-delta-rho-0}), yields 
\begin{equation}
\frac{\Phi_0}{\rho_0} \nabla^2 E_{\Phi}+
4\pi G E_{\Phi} +\frac{2}{\rho_0}
\frac{ d \Phi_0}{d r}
         \frac{d E_{\Phi}}{d r} =
4\pi G E_{\rho}.
\label{eq:relative-delta-rho-1}
\end{equation}
We are interested in the local extrema of $E_{\Phi}$. 
There are $n_{\rm max}$ number of such points whose existence 
is deduced from the oscillatory nature of basis functions. 
The derivative $d E_{\Phi}/d r$ vanishes at the extrema 
of $E_{\Phi}$ and equation (\ref{eq:relative-delta-rho-1}) 
reads 
\begin{equation}
\frac{E_{\rho}}{E_{\Phi}}=1+
\frac{\Phi_0}{4\pi G\rho_0} 
\frac{1}{E_{\Phi}}\frac{d^2 E_{\Phi}}{d r^2}.
\label{eq:relative-delta-rho-2}
\end{equation}
This is a useful relation that gives a credible estimate of 
$E_{\rho}/E_{\Phi}$ based on the quotient $\Phi_0/\rho_0$ and 
the curvature of $E_{\Phi}$. For each model, we have independently 
computed $\delta_{\rho}$ from (\ref{eq:relative-delta-rho-2}) and 
have plotted the results (scattered squares in Figure \ref{fig3}) 
against the numerical graph of $\delta_{\rho}$ obtained from the 
series expansion of $\rho_0(r)$. There is a close agreement between 
the results of two methods, confirming the fact that the drift 
$\delta_{\rho}-\delta_{\Phi}$ is independent of the choice of 
basis set and it persists in any series solution of Poisson's 
equation. 

Neither the isochrone nor Dehnen's $\gamma=1/2$ models 
match the zeroth order terms of CB73, HO92 and our basis functions. 
Therefore, the performance of these basis sets can be fairly 
compared by expanding the isochrone and Dehnen's $\gamma=1/2$ 
models (Figure \ref{fig4}). It is seen that CB73 functions have a 
poor performance in reproducing both models. Our functions have 
performed better than HO92 functions for $r\lesssim 1$ in the 
isochrone model. Nevertheless, HO92 functions have resulted in 
the lowest magnitudes of $\delta_{\Phi}$ and $\delta_{\rho}$ 
for $r\gtrsim 1$ in the isochrone model, and for $r\gtrsim 0.005$ 
in Dehnen's $\gamma=1/2$ model. Our results show that the envelopes 
of basis functions {\it must follow} the radial profiles of both 
the density and potential functions of a spherical stellar system
to assure a reliable expansion. Dehnen's shallow density cusp 
cannot be reproduced even by cuspy set of HO92 (see Figure \ref{fig4}{\em b}) 
because the central envelope of HO92's density functions is proportional 
to $r^{-1}$ while Dehnen's density profile diverges as $r^{-1/2}$. 

\begin{figure*}
\centerline{\hbox{\includegraphics[width=0.5\textwidth]
             {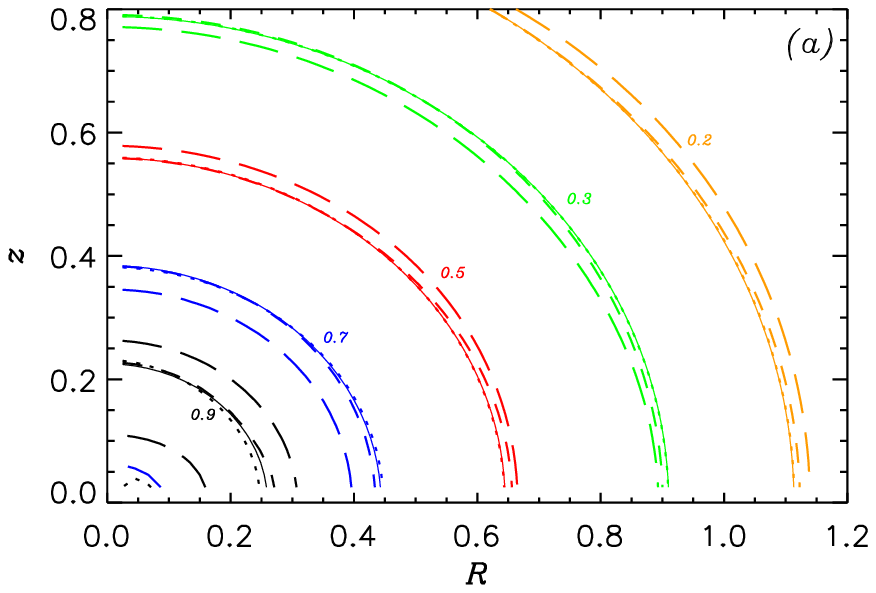}}
            \hbox{\includegraphics[width=0.5\textwidth]
             {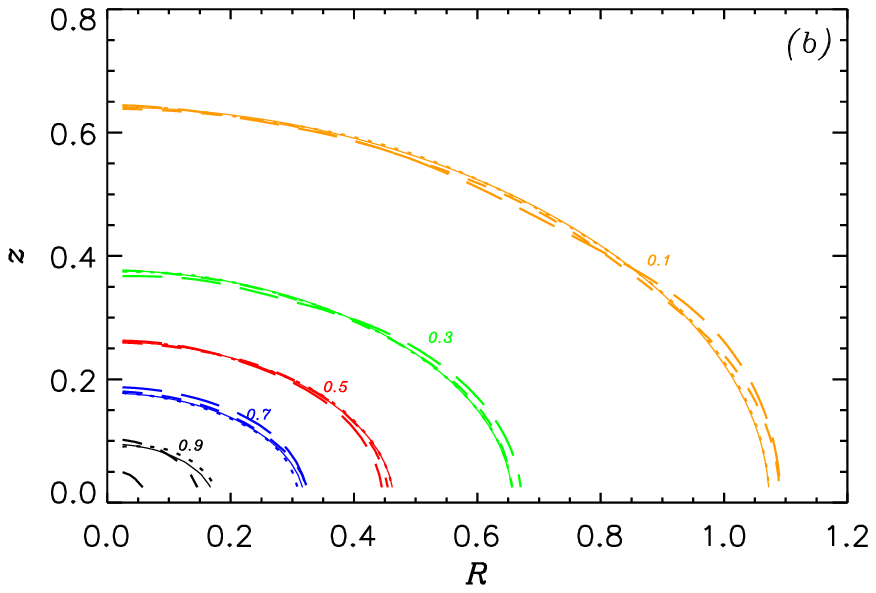}} }
\caption{{\it Left panel}: Density isocontours of a perfect 
spheroid. Solid lines correspond to the exact model density and 
short-dashed, long-dashed and dotted curves are associated with 
density expansions using our, CB73 and HO92 basis functions, respectively. 
The ellipticity of the model is $e=0.5$. {\it Right panel}: Same as the 
left panel but for an oblate Kuzmin-Kutuzov model with $c/a=0.5$. In both
figures, the levels of isocontours indicate the fraction of maximum density.}
\label{fig5} 
\end{figure*}
\begin{figure*}
\centerline{\hbox{\includegraphics[width=0.5\textwidth]
             {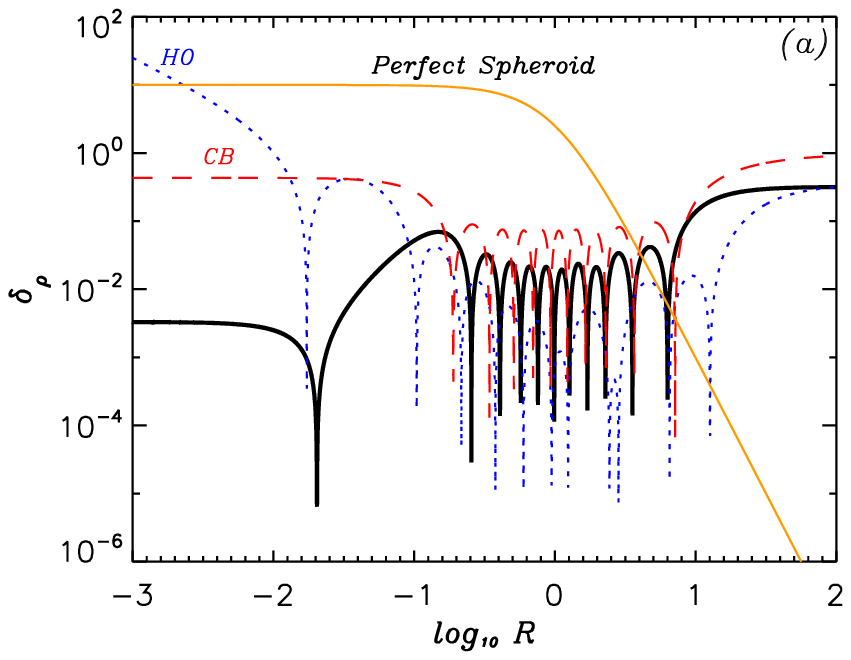}}
            \hbox{\includegraphics[width=0.5\textwidth]
             {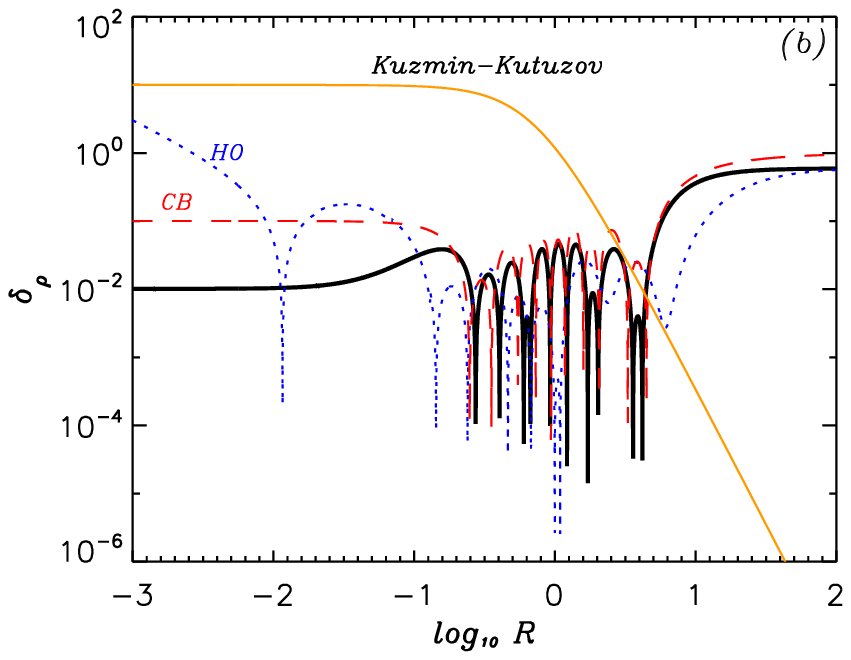}} }
\caption{Radial variation of the relative error $\delta_{\rho}$ 
(oscillatory curves) against the model density on the equatorial 
plane ($z=0$) of the same prefect spheroid (panel {\em a}) and 
Kuzmin-Kutuzov (panel {\em b}) models of Figure \ref{fig5}. 
Dashed, dotted and thick solid curves correspond, respectively, 
to CB73, HO92 and our basis functions.}
\label{fig6}
\end{figure*}

\subsection{Oblate Galaxy Models}
\label{oblate}

The modeling of oblate galaxy models is a bigger challenge because
the series of radial basis functions must converge together with 
spherical harmonics. It is therefore hard to predict how the 
combination of radial and angular functions will behave. As our 
case studies of spheroidal galaxy models, we choose an oblate 
\citet{KK62} model and a perfect spheroid \citep{dZ85}, and 
reproduce their density distributions using the series of CB73, 
HO92 and our new biorthogonal sets. In the spherical limit, the 
Kuzmin-Kutuzov model reduces to H\'enon's (1959) isochrone, and 
the perfect spheroid becomes the perfect sphere, which is the 
lowest order term of our new basis set. Since our functions 
showed slow convergence for Dehnen's spherical models near the
centre (see Figures \ref{fig2} and \ref{fig3}), we did not extend 
our analysis to their flattened \citep{DG94} counterparts. 
Moreover, we proved in \S\ref{sec:spherical-models} that the 
potential expansions are always more accurate than the density 
ones. This applies to oblate models as well, and therefore, 
we confine ourselves to computing $\delta_{\rho}$. \\
 
Defining $u^2=a^2c^2+c^2 R^2+a^2 z^2$, the density functions 
of the Kuzmin-Kutuzov and perfect spheroidal models are 
respectively given by \citep{DD88,dZ85}
\begin{subequations}
\label{ob_KK-and-PS}
\begin{align}
\rho_{\rm KK}(R,z) &= 
      \frac{Mc^2}{4\pi} \frac{(a^2-c^2) R^2 +a^4+2u^2+3a^2 u}
      {u^3\left (R^2+z^2+a^2+c^2+2u \right)^{3/2}}, \\
\rho_{\rm PS}(R,z) &= \frac{M}{\pi^2 \sqrt{1-e^2}}
     \left (1+R^2+\frac{z^2}{1-e^2} \right)^{-2},
\end{align}
\end{subequations}
where $R=r\sin\theta$ and $z=r\cos\theta$, and $z$ is the symmetry 
axis. The parameter $e$ is the flattening of the perfect spheroidal  
model. The Kuzmin-Kutuzov model has equipotential surfaces of 
the axis ratio $\sqrt{c/a}$ near the centre, and we choose its
length scale so that $a+c=1$. Here again, $M$ is the total mass 
of the galaxy. 

The isocontours of the original and expanded density functions are 
displayed in Figure \ref{fig5} for a perfect spheroidal model of 
$e=0.5$ and for a Kuzmin-Kutuzov model of $c/a=0.5$. We have set 
$r_0=1$ and $n_{\rm max}=9$, and used $l_{\rm max}=8$ for the 
Kuzmin-Kutuzov model and $l_{\rm max}=4$ for the perfect spheroid, 
respectively. The maximum deviation from the original model occurs 
near the $R$-axis because of the slow convergence of spherical 
harmonics as $\theta\rightarrow \pi/2$. Therefore, we have shown 
in Figure \ref{fig6} the variation of $\delta_{\rho}$ versus $R$ 
in the equatorial plane. Note that the existence of a symmetry 
axis implies $D_{nlm}=0$ for odd latitudinal quantum numbers and 
for $m\not =0$.

Our experiments show that by increasing $l_{\rm max}$ the density expansion 
near the equatorial plane is improved. It is evident that HO92 functions 
have failed in reproducing the finite central densities of both models 
but they have best fitted the outer parts. For $R \gtrsim 0.2$, the 
error indicator $\delta_{\rho}$ is smaller for HO92 functions than CB73 
ones by almost one order of magnitude, and that of our new functions 
lies between them. Nonetheless, only our functions result in very small 
error level of $\le 1\%$ for $R\lesssim 0.2$ in both models. This 
shows that our new basis set is the most trusted tool for modeling 
all parts of cored, oblate galaxies whose outer potential and density 
profiles fall off similar to $r^{-1}$ and $r^{-4}$, respectively. 
We note that the magnitude of $\delta_{\rho}$ rises substantially
and then saturates beyond the radial distance $R\approx 10$ where 
the density has fallen to $0.1\%$ of its central value. This property 
is shared by all tested basis sets. It is by increasing the number 
of radial basis functions ($n_{\rm max}$) together with the precision 
of computations that the error magnitude is suppressed at large radii. 

It is helpful to compare our results with \citet{Robijn96} who have 
designed a set of basis functions for the perfect spheroidal 
models. Their functions have been orthonormalised using Gram-Schmidt 
procedure. For $n_{\rm max}=9$ and $l_{\rm max}=4$ that match the number 
of series terms in our setup, they reported a maximum error of 
$\delta_{\rho} \approx 30\%$ in the density expansion for a perfect 
spheroid of ellipticity $e=0.5$ and inside the domain $0< R,z <2$. 
In the same region, our density expansion leads to a maximum error 
of $\delta_\rho \approx 7\%$, which is notably small.

\section{CONCLUSION}
\label{conclusion}

The lack of suitable PD basis sets is a serious problem in 
dynamical studies that solve Poisson's equation using series 
expansions. For three dimensional stellar systems only few 
analytic basis sets have been found and most researchers have 
tailored numerical functions to cope with their specific problems.
In this paper we generalised Clutton-Brock's (1972) idea to three 
dimensional systems and introduced a new set of basis functions, 
which have the useful property of biorthogonality. Our functions 
complement the CB73, HO92 and Zhao's (1996) basis sets because 
neither of them exhibits the following properties together: 
(i) A finite central density. (ii) An outer density fall-off 
similar to $r^{-4}$. For instance, the integrable models of 
de Zeeuw (1985) and their perturbed states, can be efficiently 
expanded by our basis functions. Thus, we get one step closer 
to the stability analysis of elliptical galaxies whose potentials 
are of St\"ackel form. \citet{Robijn95} and \citet{SV97} investigated 
the instabilities of some spheroidal galaxy models but calculating 
the eigenspectra of more general triaxial systems remains as a big 
challenge. 

Our functions were derived in terms of elementary rational, and 
associated Legendre functions for which recursive formulae are 
available. We carried out a mathematical error analysis and then 
compared its results by numerical experiments to show that density 
expansions converge slower than potential ones. By expanding 
several spherical and oblate galaxy models, we showed that an 
improper choice of basis functions can contribute potentially 
dangerous errors to dynamical studies. Not only the nature of 
the galactic centre (cuspy or cored) is an important factor 
for the selection of basis functions, the outer density and 
potential profiles also matter. Neither our new set, nor other 
basis functions cited in this paper, are suitable for the modeling 
of cuspy dark matter halos whose density profiles decay outward 
like $r^{-3}$. It is possible to find basis sets compatible with 
such systems, but that will require other choices of the weighting 
functions $g_{nl}(r)$ that must be orthogonal over the $r$-domain 
in three dimensions. It is remarked that we had set the length 
scale of our basis functions to $r_0=1$ in all of our case studies, 
but there is always an optimum value of $r_0$ that gives the best 
fit. For example, the isochrone model is best fitted by setting 
$r_0=2b$. We therefore recommend an optimal search for finding 
the best minimiser of $\delta_{\Phi}$.

\section*{Acknowledgments}
AR was supported by a Huygens Fellowship awarded by the 
Dutch Ministry of Culture, Education and Science. 
We thank the referee for a useful report.

\end{document}